\documentclass[pre,twocolumn,groupedaddress]{revtex4}
\usepackage{graphicx}
\begin{document}
\title{Exploring the origins of the power-law properties of energy landscapes: An egg-box model} 
\author{Claire P. Massen}
\affiliation{University Chemical Laboratory, Lensfield Road, Cambridge CB2 1EW, United Kingdom}
\author{Jonathan P.~K.~Doye}
\address{Physical and Theoretical Chemistry Laboratory,
Oxford University, South Parks Road, Oxford OX1 3QZ, United Kingdom}
\author{Rupert W. Nash}
\affiliation{School of Physics, University of Edinburgh, Mayfield Road, Edinburgh, EH9 3JZ, United Kingdom}
\date{\today}

\begin{abstract}

Multidimensional potential energy landscapes (PELs) have a Gaussian distribution for the energies of the minima, but at the same time the distribution of the hyperareas for the basins of attraction surrounding the minima follows a power-law.
To explore how both these features can simultaneously be true, we introduce an ``egg-box'' model.
In these model landscapes, the Gaussian energy distribution is used as a starting point and we examine whether a power-law basin area distribution can arise as a natural consequence through the swallowing up of higher-energy minima by larger low-energy basins when the variance of this Gaussian is increased sufficiently.
Although the basin area distribution is substantially broadened by this process,
it is insufficient to generate power-laws, highlighting the role played by the inhomogeneous distribution of basins in configuration space for actual PELs. 

\end{abstract}

\maketitle

\section{Introduction}

Potential energy landscapes describe how the potential energy of a system depends on the coordinates of its constituent atoms \cite{Landscapes}.
As the thermodynamics and dynamics of a system are controlled by the PEL, 
there has been much interest in exploring the insights into the behaviour of relatively complex systems that can be obtained by characterizing the features of the PEL.
Particular active research areas for this landscape approach have been supercooled liquids and protein folding.
For example, the dynamics of fragile liquids change at lower temperatures, as the system explores low-energy regions of the PEL, where the effective activation energies for structural relaxation are larger \cite{Debenedetti01}.
Furthermore, due to the astronomical number of conformations, the Levinthal paradox predicts that protein folding would be extremely unlikely to occur if these conformations were searched randomly \cite{Levinthal}.
However, a funnel on the PEL directs the protein towards its folded state \cite{Leopold,Bryngelson95}.

Here, our focus is not so much on how to relate the behaviour of the system to the underlying PEL, but rather it is to understand the fundamental properties of PELs better.
When classifying high-dimensional PELs it is common to focus on the stationary points of the PEL, particularly the minima and transition states (first-order saddle points).
Fundamental results concerning these stationary points include that the number of minima scales exponentially with the system size \cite{Stillinger82,Stillinger99,Tsai93}, and that the distribution of energies of the minima is Gaussian.
The latter has been found empirically for a range of systems \cite{Buchner99,Sciortino99}, and can be justified theoretically from the central limit theorem, assuming the system can be broken down into independent subsystems \cite{Heuer00}.

More recently, a network approach has been applied to gain insights into the connectivity of PELs \cite{Doye02, Doye05b}.
In these ``inherent structure'' networks, the minima correspond to nodes, and two nodes are joined by an edge if the corresponding minima are directly connected by the two steepest descent pathways going away from a transition state.
Such networks are dynamically motivated since if the temperature is not too high, the dynamics can be separated into vibrational motion in the basin surrounding a minimum with hopping between minima via transition state valleys on longer time scales \cite{Stillinger84,Schroder98}.

\begin{figure}
\centerline{\includegraphics[width=6.6cm]{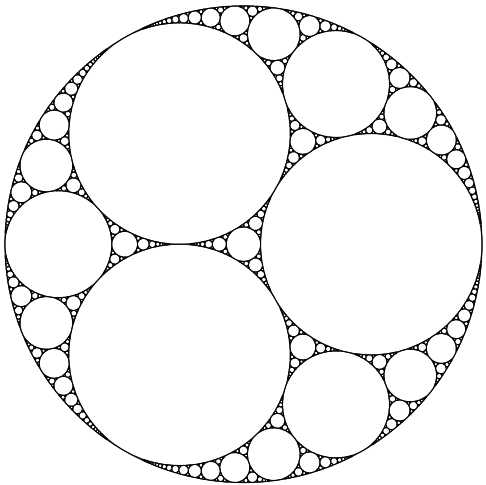}}
\caption{
An Apollonian packing.
Space is filled with different sized disks, starting from an initial configuration where the three larger disks are placed within the bounding circle.
The packing is generated iteratively, with the largest disk possible added to each gap at each iteration.
This process is continued ad infinitum, thus filling the space with successively smaller and smaller disks.
}
\label{fig:apollo}
\end{figure}

Interestingly, for small clusters, for which the complete inherent structure networks can be obtained, these networks have been found to be scale-free, that is the degree (the number of connections to a node) distribution has a power-law tail \cite{Barabasi99}.
For such a topology, most of the nodes have relatively few connections, but there are a small number of hubs that have a large number of connections, and which play an important role in connecting up the landscape.
Given that these inherent structure networks are based on the adjacency of basins in configuration space, it may seem surprising that the networks are so heterogeneous, and not more lattice-like.
However, the energy of the minima plays a key role, with the lower energy minima acting as hubs in the network, because they have larger basins of attraction, and so can have many transition states along the boundary.
Thus this leads to a picture with larger basins surrounded by smaller basins, which are in turn surrounded by smaller basins, and so on. 
Apollonian packings (see Fig.~\ref{fig:apollo} for a two-dimensional example) provide a model of how such a hierarchical arrangement can be achieved.
Furthermore, the network of contacts between the disks in such a packing is also scale-free \cite{Andrade05,Doye05}.
The analogy holds in one further way.
The distribution of disk areas in the Apollonian packing follows a power-law.
Similarly, the distribution of basin areas on a range of PELs has also been found to follow a power-law with exactly the predicted exponent \cite{Massen05b}.

Despite the success of the Apollonian analogy, one of its deficiencies is that it provides no explanation for the role of energy in the organization of the landscape.
In particular, it is not clear how a Gaussian distribution for the energies of the minima is compatible with the power-law area and degree distributions.
The broad area distribution implies that there are some very big basins and lots of small basins in the landscape, whereas a Gaussian energy distribution shows much less variation from the mean, and is symmetrical.
In order for both of these distributions to occur, the dependence of the basin area on the energy of a minimum is necessarily very steep, with the increase in the area as the energy decreases having to be faster than exponential.
The aim of this work is to explore a model landscape (which we call the ``egg-box'' model) to determine whether the broad area distribution might actually be a natural consequence of the Gaussian energy distribution, or whether it is due to a particular way that PELs are organized.
We will describe the nature and resulting properties of our model landscapes in Section \ref{sec:model}, and discuss the implications of the results in Section \ref{sec:concs}.

\section{The egg-box model}
\label{sec:model}

\begin{figure}
\centerline{\includegraphics[width=8.6cm]{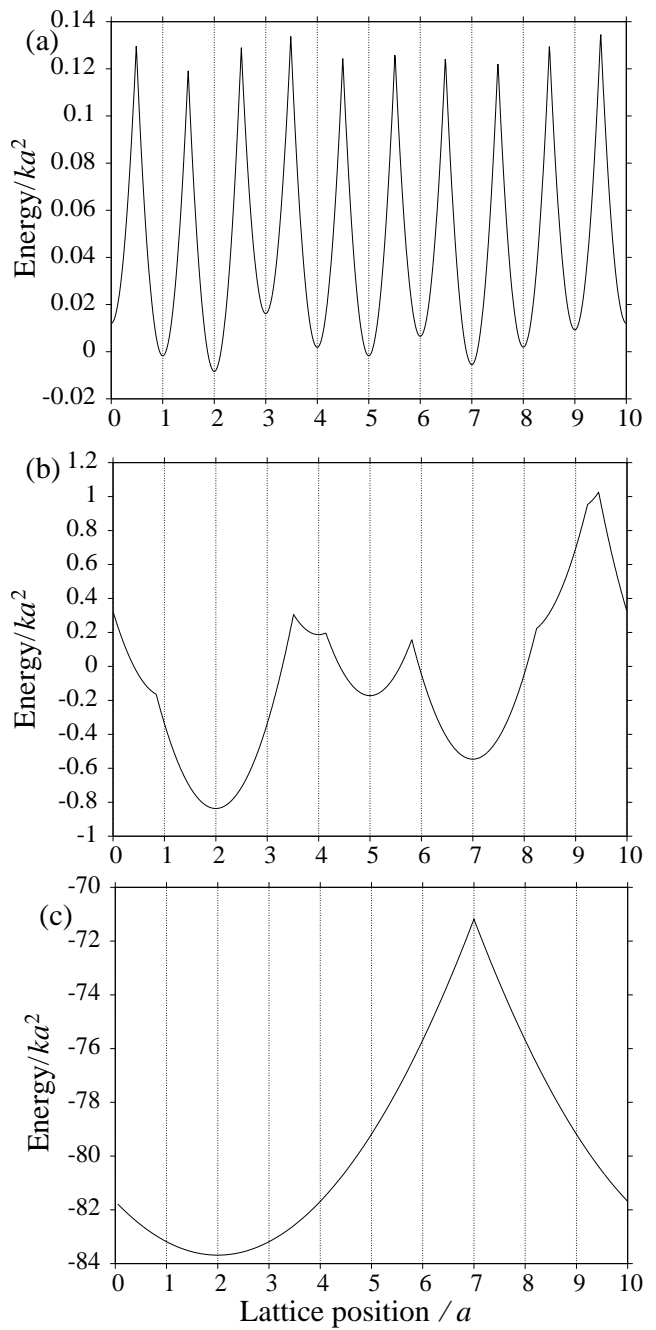}}
\caption{
Examples of a one-dimensional egg-box model with
(a) $\sigma^*=0.01$ 
(b) $\sigma^*=1$ 
(c) $\sigma^*=100$,
where $\sigma^*$ 
is the standard deviation of the Gaussian energy distribution for the minima
in reduced units.
Initially there are ten minima, but as $\sigma^*$ increases, successive minima get swallowed up, until there is only one low-energy minimum remaining.
}
\label{fig:sigma}
\end{figure}

The idea of our model is to assign minima (which are surrounded by harmonic basins) to particular positions in configuration space, typically a lattice, e.g.~a square lattice in two dimensions.
If these minima all have the same energy, then each basin of attraction will have the same area, and the same connectivity.
This, if you like, is the ``egg-box'' limit.
However, if instead the energies of the minima have a Gaussian distribution of sufficient variance, there is the possibility that some of the higher-energy minima will get swallowed up by the basins of lower energy minima, as illustrated in Fig.~\ref{fig:sigma}.
This process will inevitably lead to a broader area distribution and, for dimension greater than one, a broader degree distribution.
But, the question is whether it will be sufficient to generate the power-law area and degree distributions that have been empirically observed for real PELs.

\begin{figure}
\centerline{\includegraphics[width=8.6cm]{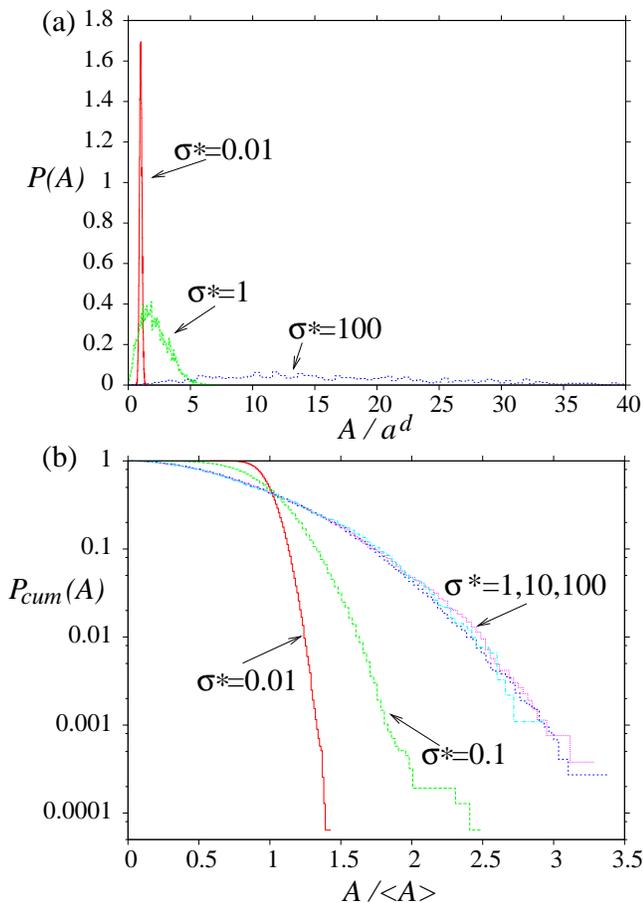}}
\caption{(Colour online)
Variation of the basin area distribution with the width of the Gaussian energy distribution for the one-dimensional egg-box model.
(a) Area distributions, (b) cumulative distributions, with the area measured with respect to its mean.
Each is for the one dimensional system with 15\,625 minima initially (some may be eaten).
}
\label{fig:sigma-pa}
\end{figure}

More formally, the potential energy for our model is given by 
\begin{equation}
V({\mathbf r})=\min\left[\left\{V_i({\mathbf r})\right\}\right],
\end{equation}
where 
\begin{equation}
V_i({\mathbf r})=E_i+{1\over 2} k \left|{\mathbf r}-{\mathbf r_i}\right|^2,
\end{equation}
${\mathbf r_i}$ is the site of a harmonic potential and $E_i$ is the energy at 
its bottom. Each site will give rise to a minimum on the PEL, except for those 
sites where $V({\mathbf r_i})<E_i$. 
For the sites of the initial minima we use a hypercubic lattice, and apply periodic boundary conditions.
The parameters of this egg-box model are the lattice spacing $a$, the force constant $k$ of the harmonic potential surrounding each minimum, and the mean $E_0$ and standard deviation $\sigma$ of the Gaussian.
We always centre the Gaussian at zero, leaving three remaining variables, two of which define the units of energy and length, with one effective parameter.
We use $a$ and $ka^2$ as the units of length and energy respectively, and choose to vary $\sigma^*=\sigma /ka^2$.
As already mentioned, as $\sigma^*$ is increased the landscape becomes more heterogeneous (Fig.~\ref{fig:sigma}).
One obvious question is whether this evolution reaches some kind of steady state.
The results in Fig.~\ref{fig:sigma-pa} for a one-dimensional example show that although the mean basin area will continue to increase with $\sigma^*$, the distribution tends to a limiting form.
We focus on the landscapes produced in this limit since this is when the area distributions will be at their broadest.
Fig.~\ref{fig:sigma-pa} suggests that landscapes with $\sigma^* = 1$ will be sufficient for this purpose, and this is the value we generally use.

\begin{figure}
\centerline{\includegraphics[width=7.6cm]{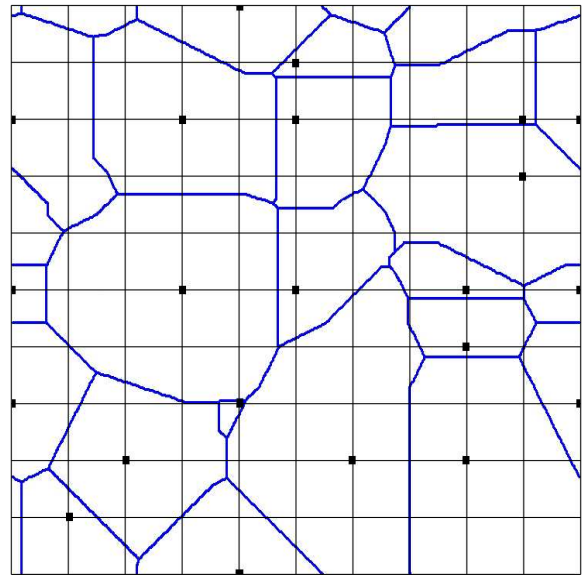}}
\caption{
Example of the egg-box model in two dimensions.
The grid shows the positions of the initial minima, 
the black squares represent those that remain at $\sigma^*=1$, 
and the thick lines represent the boundaries of their basins of attraction.
}
\label{fig:2dpic}
\end{figure}

In order to compare to the egg-box model to PELs, we of course need to examine the model in higher dimension.
Some of these effects are illustrated in the two-dimensional example shown in 
Fig.~\ref{fig:2dpic}.
Firstly, the basins exhibit a wider range of areas.
However, the area distribution is clearly much narrower than that for the Apollonian packing in Fig.~\ref{fig:apollo}, and there is no sign of any hierarchical ordering.
Secondly, as the dimension is increased, more basins are eaten because each 
basin has more neighbours and the largest possible degree increases.

\begin{figure}
\centerline{\includegraphics[width=8.6cm]{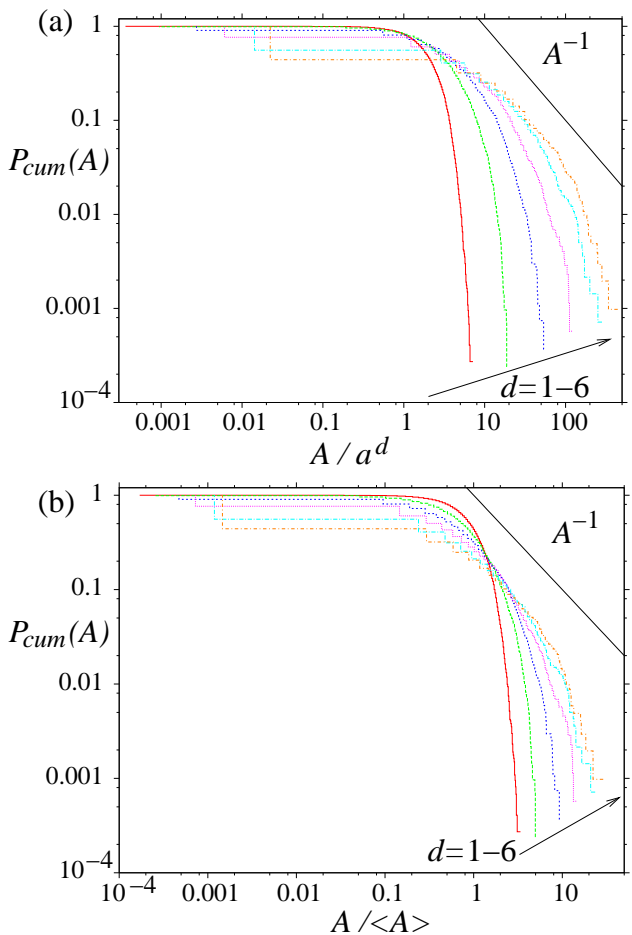}}
\caption{(Colour online)
Variation of the cumulative basin area distribution with the 
dimension of configuration space for $d=1$--6.
In (a) the distribution is plotted against the absolute area, whereas in (b) the area is measured relative to its mean value.
Also shown is the power-law $P_{cum}(A) \sim A^{-1}$ which is the form seen for high-dimensional Apollonian packings \cite{Doye05} and various PELs \cite{Massen05b}.
}
\label{fig:dim-pa}
\end{figure}

Quantitative results have been obtained for systems with up to six dimensions and around 15\,625 initial minima.
For higher dimensions, analytical approaches to determine the shapes of the basins, and hence their areas, become increasingly involved, so we instead use a Monte Carlo approach.
A large number of points were chosen at random and the basins to which the points belong were determined.
The area of a basin is then proportional to the number of points assigned to that minimum.
Area distributions are shown in Fig.~\ref{fig:dim-pa} for increasing dimension.
There is a wider range of basin sizes in higher dimension, where small changes in radius have a larger effect.
The average and maximum areas also increase because more basins are eaten.

\begin{figure}
\centerline{\includegraphics[width=8.6cm]{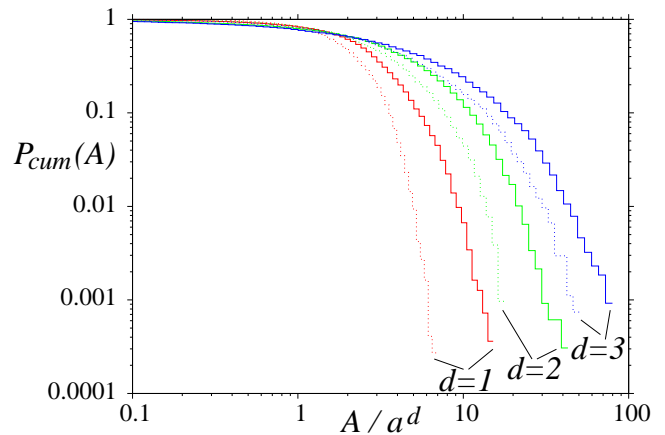}}
\caption{(Colour online)
Effect of the positions of the initial minima on the basin area distribution 
for 1--3 dimensions.
Dashed lines are for when the initial minima are on a lattice 
(as for Fig.~\ref{fig:dim-pa}(a)) and 
solid lines are for when they are distributed randomly.
}
\label{fig:lat-pa}
\end{figure}

In the results shown so far, the minima have initially been placed on a lattice.
We also consider landscapes where the initial minima have instead been given random positions.
As shown in Fig.~\ref{fig:lat-pa}, this broadens the area distribution because the starting points are less uniformly separated.
For example, when $\sigma^*=0$, the area distribution is a delta function if the basins start on a hypercubic lattice.
However, there will be some heterogeneity if the basins are given random starting points, leading to a broader distribution. 
This effect is weaker in higher dimensions, 
and so the choice of the initial sites becomes less important. 

We can compare the area distributions seen for the egg-box model to that seen for PELs \cite{Massen05b} and for the Apollonian packing \cite{Doye05}.
The latter two both follow a power-law with exponent close to $-2$ (or $-1$ for the cumulative distribution), and such a power-law has also been plotted in Fig.~\ref{fig:dim-pa} for comparison.
This form is tangential to the observed basin area distributions, but only for a small range of areas, and the distributions show significant curvature.
By contrast, the area distributions for PELs associated with common model liquids followed an approximate power-law over up to 18 decades.
However, the configuration space for the latter had 765 dimensions.
Therefore, it is natural to ask whether in the limit of much higher dimensions the egg-box model might show similar behaviour.

\begin{figure}
\centerline{\includegraphics[width=8.6cm]{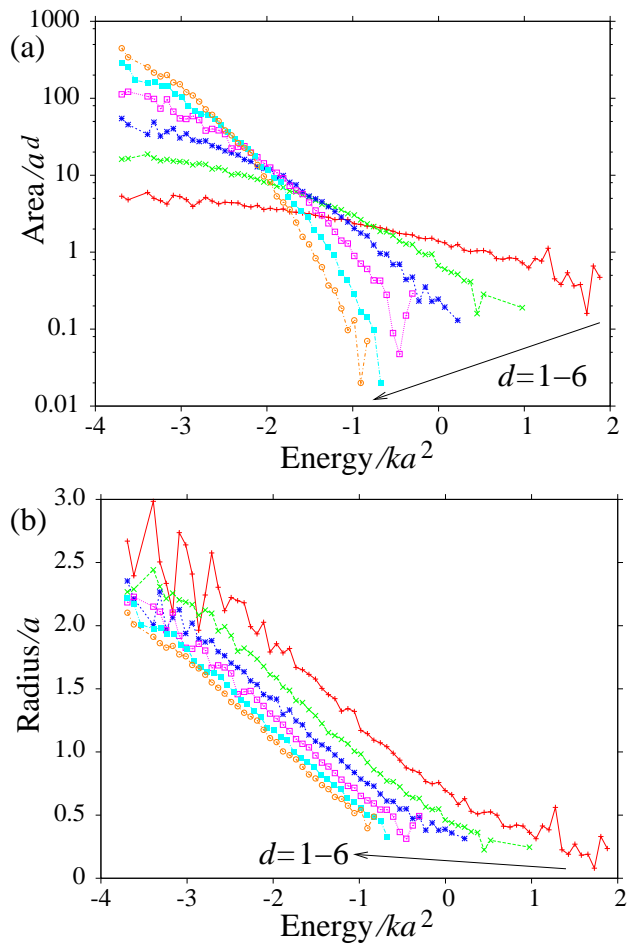}}
\caption{(Colour online)
(a) Basin area and 
(b) average radius as a function of energy for dimensions 1--6.
Each point represents an average of those minima in a given energy bin.
}
\label{fig:ae}
\end{figure}

Extending our results to significantly higher dimensions, however, is not feasible because of the rapidly increasing number of lattice points required.
Instead, our approach is to see whether the trends with increasing dimension are in the right direction to make a power-law a possibility.
In this regard, the variation of basin area with energy of the minimum is particularly informative (Fig.~\ref{fig:ae}).
As expected, low-energy basins are bigger.
Furthermore, the energy-dependence of the basin area becomes steeper with increasing dimension, because the maximum area increases, giving a wider range of areas, and $E_{max}$ decreases as high-energy basins are more likely to be eaten.
Furthermore, small changes in radius have a larger effect due to the higher dimension.

Most importantly, the curvature of $\log A(E)$ is negative.
This contrasts with the PELs studied, which showed the positive curvature that is necessary to give rise to a power-law area distribution.
In the egg-box model, as we go to lower energies, the curve gets flatter.
Therefore, we have to go to much lower energies to see basins with big areas, but a Gaussian energy distribution means that very low-energy minima are very rare.
Furthermore, at high energy, there is a steeper cut-off for the egg-box model, meaning there are fewer small basins.
This is in the region of the tail of the Gaussian distribution, where $A(E)$ would need to be shallow in order to see lots of high-energy, low-area basins, as is seen in PELs.

These trends suggest that, with increasing dimension, the egg-box model will not become more like PELs.
On the contrary, the negative curvature becomes even more pronounced.
This is confirmed on examining the dependence of the radius on energy in Fig.~\ref{fig:ae}(b), where the radius of a basin was calculated from the area by assuming that the basin was hyperspherical.
$r(E)$ is almost linear, and seems to be tending towards a limiting form, thus further implying that the curvature of $\log A(E)$ will be even more negative in higher dimension.

\begin{figure}
\centerline{\includegraphics[width=8.6cm]{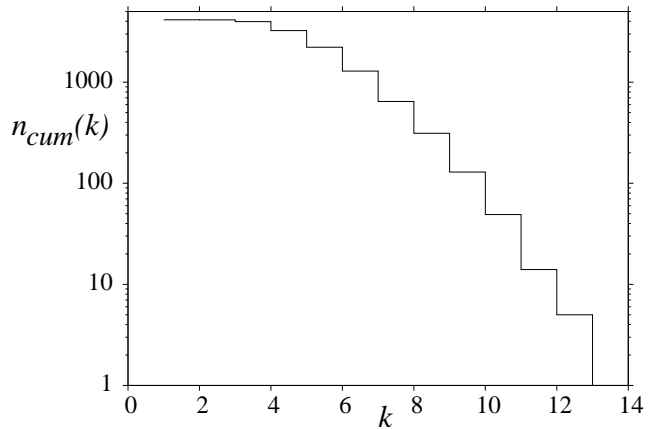}}
\caption{
Cumulative degree distribution for the two-dimensional egg-box model with
$\sigma^*=1$.
}
\label{fig:pk}
\end{figure}

Some network properties for the two-dimensional egg-box have also been studied.
Initially, there were 15\,625 minima.
As some of these are eaten, the final number of minima is $N=4140$, so the size of the network is very similar to that of the ISN for the 14-atom cluster LJ$_{14}$ studied previously \cite{Doye02}.
The number of edges in the network is $M=12\,142$, giving an average number of connections per node (average degree) of $\langle k \rangle = 2M/N = 5.87$.
This can be directly compared to the ISN of LJ$_{14}$, which has $\langle k \rangle = 29.12$.
Therefore, the egg-box model has much lower overall connectivity than the PEL.
The maximum connectivity, $k_{max}=13$, is again much lower than that seen for LJ$_{14}$, where the highest-connected basin is connected to 3201 others.

The degree distribution for the 2D egg-box model is roughly exponential, as shown in Fig.~\ref{fig:pk}.
By contrast, the networks associated with PELs and the Apollonian packing have scale-free character, i.e.~their degree distributions follow power-laws.
This behaviour is due to the presence of some very large basins, with very long boundaries, which have high degree and act as hubs, connecting up the networks.
By contrast, the egg-box model features neither very large basins to act as hubs, nor many small basins around their boundaries, as evident from comparing Figs.~\ref{fig:apollo} and \ref{fig:2dpic}.

The network of the egg-box model has very long paths between nodes.
This result cannot be compared directly to the PELs due to the different sizes and connectivity patterns, both of which affect path lengths.
However, if we compare the path length of the egg-box to that of a random network of the same size and having the same degree distribution, we see that the average path length is roughly four times that of the random network.
By contrast, for the PEL of clusters LJ$_{7}$ to LJ$_{14}$, the path length is between 1-1.05 times that for a random network.
The egg-box model therefore has very long path lengths.
The egg-box model is also very highly clustered, i.e.~the neighbours of a node are also likely to be connected.
Its clustering coefficient \cite{Watts98} is approximately 30 times that for a random network, compared to roughly 1.6 times for the ISN of LJ$_{14}$.
Therefore, with its long path lengths and high clustering, the egg-box model has very similar properties to a lattice.
Despite the broadening of the area distribution resulting from the swallowing up of higher-energy minima as $\sigma^*$ is increased, the contact network between the basins does not change its essential character.
Although it is possible that these properties may change for higher dimensions, it has been shown that PELs behave differently to high-dimensional lattices, in particular having shorter path lengths \cite{Doye02,Doye05b}.

\section{Conclusions}
\label{sec:concs}

The aim of this paper was to determine whether the power-law area distribution observed in PELs is a natural consequence of a Gaussian distribution of energies for the minima, or whether it reflects further ordering of the PEL.
We do see some of the expected trends.
Increasing the width of the Gaussian leads to a broadening of the area distribution, where lower-energy basins have higher area and higher degree.
However, this is not sufficient to reproduce the properties of real PELs.
In some ways, this might be regarded as a negative result, however, it allows us to get a better understanding of what the key features of real PELs are that lead to the observed properties.

Some assumptions are used in the egg-box model that constitute possible causes of the differences between the resultant model landscapes and PELs.
Firstly, the dimension of the models studied is much lower than that for PELs of all but the smallest systems.
However, the trends for the dependence of the area distribution on dimension that we observed are not going in the right direction.
Thus higher-dimensional model landscapes are unlikely to correct the deficiencies of the lower-dimensional models examined here.

Secondly, we assumed that the basins have a harmonic form.
Generally, it is found that applying the harmonic approximation to the basins of real PELs gives a correct qualitative description of the systems' behaviour \cite{Doye95,Sciortino00}, and so we think this assumption is unlikely to have a significant effect on our conclusions.
Thirdly, we assumed that the force constant $k$ was the same for all basins, independent of the basin energy.
However, given that the vibrational frequencies for clusters \cite{Wales93,Noya06} and for liquids at constant volume \cite{Middleton01b}
have an opposite dependence on energy, yet the PELs of examples of both of 
these types of system have been seen to have broad area distributions \cite{Massen05b}, we think this assumption is unlikely to have too much effect.

Fourthly, there is an absence of correlations between neighbouring minima in the egg-box model, i.e. the assigned energy of an initial minimum is independent of that of its neighbours.
However in real landscapes, the energies of neighbouring minima are correlated with minima more likely to be connected to others of similar energy \cite{Doye05b}.
The likely effect of such correlations on the current model would be to increase the number of high-energy minima as they are less likely to be swallowed up if they are surrounded by minima of similar energy, and to decrease the area of lower-energy basins as they are more likely to be surrounded by other lower-energy minima restricting the growth of their basins.
Thus the dependence of the basin area on energy would be likely to be less steep, and in itself the introduction of correlations into our model is unlikely to lead to a further broadening of the area distribution.

Fifthly, there is the homogeneous distribution of initial minima in the egg-box model.
Both the lattice and the random initial configurations lead to an essentially uniform density of points, in contrast with the Apollonian packing which is very heterogeneous, with small basins clustered in the gaps between larger basins.
Such an inhomogeneous distribution of minima combined with energy correlations, where the high-energy minima are more closely spaced and the low-energy minima are more separated, would potentially lead to larger areas for the lower-energy minima and a greater number of small high-energy minima, thus impacting on the curvature of $A(E)$ and broadening the area distribution.

Thus, the differing properties of our egg-box model landscapes and real PELs are most likely due to the inhomogeneous distribution of minima in configuration space, and the hierarchical nature of the packing of the basins of attraction on the PEL.
This conclusion therefore highlights the role these features play in giving rise to the power-law basin area distribution observed.

But, this of course, begs the question what then are the sources of the inhomogeneity.
This will be the subject of future work, but some partial answers come from a further difference between the egg-box model and real PELs, namely the lack of a corresponding real space atomic structure in the egg-box model.
For example, high-energy minima on a real PEL will correspond to disordered configurations, where the structure is frustrated.
It has been found that such structures have many non-diffusive modes, which have small barriers leading to different minima, but involve no real reordering of atom positions \cite{Middleton01b}.
Thus there will be a high density of high-energy minima with small basins clustered together.
By contrast, for lower-energy minima the local environment around each atom is likely to be much closer to optimal, and to go between basins requires following a diffusive mode where there is a significant reordering of the atomic positions, associated with some atoms escaping from their caging environments.
Thus, the organization of the PEL and the atomic structure are intimately related.

\begin{acknowledgments}
The authors would like to thank the Royal Society (JPKD) and the EPSRC (CPM) for financial support.
\end{acknowledgments}


\begin{thebibliography}{25}
\expandafter\ifx\csname natexlab\endcsname\relax\def\natexlab#1{#1}\fi
\expandafter\ifx\csname bibnamefont\endcsname\relax
  \def\bibnamefont#1{#1}\fi
\expandafter\ifx\csname bibfnamefont\endcsname\relax
  \def\bibfnamefont#1{#1}\fi
\expandafter\ifx\csname citenamefont\endcsname\relax
  \def\citenamefont#1{#1}\fi
\expandafter\ifx\csname url\endcsname\relax
  \def\url#1{\texttt{#1}}\fi
\expandafter\ifx\csname urlprefix\endcsname\relax\def\urlprefix{URL }\fi
\providecommand{\bibinfo}[2]{#2}
\providecommand{\eprint}[2][]{\url{#2}}

\bibitem[{\citenamefont{Wales}(2004)}]{Landscapes}
\bibinfo{author}{\bibfnamefont{D.~J.} \bibnamefont{Wales}},
  \emph{\bibinfo{title}{Energy Landscapes}} (\bibinfo{publisher}{Cambridge
  University Press}, \bibinfo{address}{Cambridge}, \bibinfo{year}{2004}).

\bibitem[{\citenamefont{Debenedetti and Stillinger}(2001)}]{Debenedetti01}
\bibinfo{author}{\bibfnamefont{P.~G.} \bibnamefont{Debenedetti}}
  \bibnamefont{and} \bibinfo{author}{\bibfnamefont{F.~H.}
  \bibnamefont{Stillinger}}, \bibinfo{journal}{Nature}
  \textbf{\bibinfo{volume}{410}}, \bibinfo{pages}{259} (\bibinfo{year}{2001}).

\bibitem[{\citenamefont{Levinthal}(1969)}]{Levinthal}
\bibinfo{author}{\bibfnamefont{C.}~\bibnamefont{Levinthal}}, in
  \emph{\bibinfo{booktitle}{M\"ossbauer Spectroscopy in Biological Systems:
  Proceedings of a Meeting Held at Allerton House, Monticello, Illinois}},
  edited by \bibinfo{editor}{\bibfnamefont{P.}~\bibnamefont{DeBrunner}},
  \bibinfo{editor}{\bibfnamefont{J.}~\bibnamefont{Tsibris}}, \bibnamefont{and}
  \bibinfo{editor}{\bibfnamefont{E.}~\bibnamefont{Munck}}
  (\bibinfo{publisher}{University of Illinois Press},
  \bibinfo{address}{Urbana}, \bibinfo{year}{1969}).

\bibitem[{\citenamefont{Leopold et~al.}(1992)\citenamefont{Leopold, Montal, and
  Onuchic}}]{Leopold}
\bibinfo{author}{\bibfnamefont{P.~E.} \bibnamefont{Leopold}},
  \bibinfo{author}{\bibfnamefont{M.}~\bibnamefont{Montal}}, \bibnamefont{and}
  \bibinfo{author}{\bibfnamefont{J.~N.} \bibnamefont{Onuchic}},
  \bibinfo{journal}{Proc. Natl. Acad. Sci. U.S.A.}
  \textbf{\bibinfo{volume}{89}}, \bibinfo{pages}{8721} (\bibinfo{year}{1992}).

\bibitem[{\citenamefont{Bryngelson et~al.}(1995)\citenamefont{Bryngelson,
  Onuchic, Socci, and Wolynes}}]{Bryngelson95}
\bibinfo{author}{\bibfnamefont{J.~D.} \bibnamefont{Bryngelson}},
  \bibinfo{author}{\bibfnamefont{J.~N.} \bibnamefont{Onuchic}},
  \bibinfo{author}{\bibfnamefont{N.~D.} \bibnamefont{Socci}}, \bibnamefont{and}
  \bibinfo{author}{\bibfnamefont{P.~G.} \bibnamefont{Wolynes}},
  \bibinfo{journal}{Proteins} \textbf{\bibinfo{volume}{21}},
  \bibinfo{pages}{167} (\bibinfo{year}{1995}).

\bibitem[{\citenamefont{Stillinger and Weber}(1982)}]{Stillinger82}
\bibinfo{author}{\bibfnamefont{F.~H.} \bibnamefont{Stillinger}}
  \bibnamefont{and} \bibinfo{author}{\bibfnamefont{T.~A.} \bibnamefont{Weber}},
  \bibinfo{journal}{Phys. Rev. A} \textbf{\bibinfo{volume}{25}},
  \bibinfo{pages}{978} (\bibinfo{year}{1982}).

\bibitem[{\citenamefont{Stillinger}(1999)}]{Stillinger99}
\bibinfo{author}{\bibfnamefont{F.~H.} \bibnamefont{Stillinger}},
  \bibinfo{journal}{Phys. Rev. E} \textbf{\bibinfo{volume}{59}},
  \bibinfo{pages}{48} (\bibinfo{year}{1999}).

\bibitem[{\citenamefont{Tsai and Jordan}(1993)}]{Tsai93}
\bibinfo{author}{\bibfnamefont{C.~J.} \bibnamefont{Tsai}} \bibnamefont{and}
  \bibinfo{author}{\bibfnamefont{K.~D.} \bibnamefont{Jordan}},
  \bibinfo{journal}{J. Phys. Chem.} \textbf{\bibinfo{volume}{97}},
  \bibinfo{pages}{11227} (\bibinfo{year}{1993}).

\bibitem[{\citenamefont{B\"{u}chner and Heuer}(1999)}]{Buchner99}
\bibinfo{author}{\bibfnamefont{S.}~\bibnamefont{B\"{u}chner}} \bibnamefont{and}
  \bibinfo{author}{\bibfnamefont{A.}~\bibnamefont{Heuer}},
  \bibinfo{journal}{Phys. Rev. E} \textbf{\bibinfo{volume}{60}},
  \bibinfo{pages}{6507} (\bibinfo{year}{1999}).

\bibitem[{\citenamefont{Sciortino et~al.}(1999)\citenamefont{Sciortino, Kob,
  and Tartaglia}}]{Sciortino99}
\bibinfo{author}{\bibfnamefont{F.}~\bibnamefont{Sciortino}},
  \bibinfo{author}{\bibfnamefont{W.}~\bibnamefont{Kob}}, \bibnamefont{and}
  \bibinfo{author}{\bibfnamefont{P.}~\bibnamefont{Tartaglia}},
  \bibinfo{journal}{Phys. Rev. Lett.} \textbf{\bibinfo{volume}{83}},
  \bibinfo{pages}{3214} (\bibinfo{year}{1999}).

\bibitem[{\citenamefont{Heuer and B\"{u}chner}(2000)}]{Heuer00}
\bibinfo{author}{\bibfnamefont{A.}~\bibnamefont{Heuer}} \bibnamefont{and}
  \bibinfo{author}{\bibfnamefont{S.}~\bibnamefont{B\"{u}chner}},
  \bibinfo{journal}{J. Phys. Cond. Mat.} \textbf{\bibinfo{volume}{12}},
  \bibinfo{pages}{6535} (\bibinfo{year}{2000}).

\bibitem[{\citenamefont{Doye}(2002)}]{Doye02}
\bibinfo{author}{\bibfnamefont{J.~P.~K.} \bibnamefont{Doye}},
  \bibinfo{journal}{Phys. Rev. Lett.} \textbf{\bibinfo{volume}{88}},
  \bibinfo{pages}{238701} (\bibinfo{year}{2002}).

\bibitem[{\citenamefont{Doye and Massen}(2005{\natexlab{a}})}]{Doye05b}
\bibinfo{author}{\bibfnamefont{J.~P.~K.} \bibnamefont{Doye}} \bibnamefont{and}
  \bibinfo{author}{\bibfnamefont{C.~P.} \bibnamefont{Massen}},
  \bibinfo{journal}{J. Chem. Phys.} \textbf{\bibinfo{volume}{122}},
  \bibinfo{pages}{084105} (\bibinfo{year}{2005}{\natexlab{a}}).

\bibitem[{\citenamefont{Stillinger and Weber}(1984)}]{Stillinger84}
\bibinfo{author}{\bibfnamefont{F.~H.} \bibnamefont{Stillinger}}
  \bibnamefont{and} \bibinfo{author}{\bibfnamefont{T.~A.} \bibnamefont{Weber}},
  \bibinfo{journal}{Science} \textbf{\bibinfo{volume}{225}},
  \bibinfo{pages}{983} (\bibinfo{year}{1984}).

\bibitem[{\citenamefont{Schr{\o}der and Dyre}(1998)}]{Schroder98}
\bibinfo{author}{\bibfnamefont{T.~B.} \bibnamefont{Schr{\o}der}}
  \bibnamefont{and} \bibinfo{author}{\bibfnamefont{J.~C.} \bibnamefont{Dyre}},
  \bibinfo{journal}{J. Non-crystalline Solids} \textbf{\bibinfo{volume}{235}},
  \bibinfo{pages}{331} (\bibinfo{year}{1998}).

\bibitem[{\citenamefont{Barab\'{a}si and Albert}(1999)}]{Barabasi99}
\bibinfo{author}{\bibfnamefont{A.-L.} \bibnamefont{Barab\'{a}si}}
  \bibnamefont{and} \bibinfo{author}{\bibfnamefont{R.}~\bibnamefont{Albert}},
  \bibinfo{journal}{Science} \textbf{\bibinfo{volume}{286}},
  \bibinfo{pages}{509} (\bibinfo{year}{1999}).

\bibitem[{\citenamefont{Doye and Massen}(2005{\natexlab{b}})}]{Doye05}
\bibinfo{author}{\bibfnamefont{J.~P.~K.} \bibnamefont{Doye}} \bibnamefont{and}
  \bibinfo{author}{\bibfnamefont{C.~P.} \bibnamefont{Massen}},
  \bibinfo{journal}{Phys. Rev. E} \textbf{\bibinfo{volume}{71}},
  \bibinfo{pages}{016128} (\bibinfo{year}{2005}{\natexlab{b}}).

\bibitem[{\citenamefont{Andrade~Jr. et~al.}(2005)\citenamefont{Andrade~Jr.,
  Herrmann, Andrade, and da~Silva}}]{Andrade05}
\bibinfo{author}{\bibfnamefont{J.~S.} \bibnamefont{Andrade~Jr.}},
  \bibinfo{author}{\bibfnamefont{H.~J.} \bibnamefont{Herrmann}},
  \bibinfo{author}{\bibfnamefont{R.~F.~S.} \bibnamefont{Andrade}},
  \bibnamefont{and} \bibinfo{author}{\bibfnamefont{L.~R.}
  \bibnamefont{da~Silva}}, \bibinfo{journal}{Phys. Rev. Lett.}
  \textbf{\bibinfo{volume}{94}}, \bibinfo{pages}{018702}
  (\bibinfo{year}{2005}).

\bibitem[{\citenamefont{Massen and Doye}(e-print)}]{Massen05b}
\bibinfo{author}{\bibfnamefont{C.~P.} \bibnamefont{Massen}} \bibnamefont{and}
  \bibinfo{author}{\bibfnamefont{J.~P.~K.} \bibnamefont{Doye}},
  \bibinfo{journal}{cond-mat/0509185}  (\bibinfo{year}{e-print}).

\bibitem[{\citenamefont{Watts and Strogatz}(1998)}]{Watts98}
\bibinfo{author}{\bibfnamefont{D.~J.} \bibnamefont{Watts}} \bibnamefont{and}
  \bibinfo{author}{\bibfnamefont{S.~H.} \bibnamefont{Strogatz}},
  \bibinfo{journal}{Nature} \textbf{\bibinfo{volume}{393}},
  \bibinfo{pages}{440} (\bibinfo{year}{1998}).

\bibitem[{\citenamefont{Doye and Wales}(1995)}]{Doye95}
\bibinfo{author}{\bibfnamefont{J.~P.~K.} \bibnamefont{Doye}} \bibnamefont{and}
  \bibinfo{author}{\bibfnamefont{D.~J.} \bibnamefont{Wales}},
  \bibinfo{journal}{J. Chem. Phys.} \textbf{\bibinfo{volume}{102}},
  \bibinfo{pages}{9659} (\bibinfo{year}{1995}).

\bibitem[{\citenamefont{Sciortino et~al.}(2000)\citenamefont{Sciortino, Kob,
  and Tartaglia}}]{Sciortino00}
\bibinfo{author}{\bibfnamefont{F.}~\bibnamefont{Sciortino}},
  \bibinfo{author}{\bibfnamefont{W.}~\bibnamefont{Kob}}, \bibnamefont{and}
  \bibinfo{author}{\bibfnamefont{P.}~\bibnamefont{Tartaglia}},
  \bibinfo{journal}{J. Phys.: Condens. Matter} \textbf{\bibinfo{volume}{12}},
  \bibinfo{pages}{6525} (\bibinfo{year}{2000}).

\bibitem[{\citenamefont{Wales}(1993)}]{Wales93}
\bibinfo{author}{\bibfnamefont{D.~J.} \bibnamefont{Wales}},
  \bibinfo{journal}{Mol. Phys.} \textbf{\bibinfo{volume}{78}},
  \bibinfo{pages}{151} (\bibinfo{year}{1993}).

\bibitem[{\citenamefont{Noya and Doye}(2006)}]{Noya06}
\bibinfo{author}{\bibfnamefont{E.~G.} \bibnamefont{Noya}} \bibnamefont{and}
  \bibinfo{author}{\bibfnamefont{J.~P.~K.} \bibnamefont{Doye}},
  \bibinfo{journal}{J. Chem. Phys.} \textbf{\bibinfo{volume}{124}},
  \bibinfo{pages}{104503} (\bibinfo{year}{2006}).

\bibitem[{\citenamefont{Middleton and Wales}(2001)}]{Middleton01b}
\bibinfo{author}{\bibfnamefont{T.~F.} \bibnamefont{Middleton}}
  \bibnamefont{and} \bibinfo{author}{\bibfnamefont{D.~J.} \bibnamefont{Wales}},
  \bibinfo{journal}{Phys. Rev. B} \textbf{\bibinfo{volume}{64}},
  \bibinfo{pages}{024205} (\bibinfo{year}{2001}).

\end{thebibliography}
\end{document}